\documentclass[prl,english,preprintnumbers,amsmath,amssymb,nofootinbib,twocolumn,superscriptaddress]{revtex4-1}

\pdfoutput=1

\usepackage[latin1]{inputenc}
\usepackage{graphicx}
\usepackage{bbm}
\usepackage{amssymb}
\usepackage{amsmath}
\usepackage{tabularx}

\usepackage{dsfont}

\def\0#1#2{\frac{#1}{#2}}

\def\s0#1#2{\mbox{\small{$ \frac{#1}{#2} $}}}

\newcommand{\be}{\begin{eqnarray}}
\newcommand{\ee}{\end{eqnarray}}

\newcommand{\nn}{\nonumber }

\newcommand{\beq}{\begin{equation}}
\newcommand{\eeq}{\end{equation}}
\newcommand{\bea}{\begin{eqnarray}}
\newcommand{\eea}{\end{eqnarray}}

\usepackage{babel}
\makeatother
\begin{document}

\title{Zero-temperature equation of state of mass-imbalanced resonant Fermi gases}

\author{Jens Braun} 
\affiliation{Institut f\"ur Kernphysik (Theoriezentrum), Technische Universit\"at Darmstadt, 
D-64289 Darmstadt, Germany}
\affiliation{ExtreMe Matter Institute EMMI, GSI, Planckstra{\ss}e 1, D-64291 Darmstadt, Germany}
\author{Joaqu\'{\i}n E. Drut} 
\affiliation{Institut f\"ur Kernphysik (Theoriezentrum), Technische Universit\"at Darmstadt, 
D-64289 Darmstadt, Germany}
\affiliation{ExtreMe Matter Institute EMMI, GSI, Planckstra{\ss}e 1, D-64291 Darmstadt, Germany}
\affiliation{Department of Physics and Astronomy, University of North Carolina, Chapel Hill, NC 27599, USA}
\author{Dietrich Roscher}
\affiliation{Institut f\"ur Kernphysik (Theoriezentrum), Technische Universit\"at Darmstadt, 
D-64289 Darmstadt, Germany}

\begin{abstract}
We calculate the zero-temperature equation of state of mass-imbalanced resonant Fermi gases in an {\it ab initio} fashion, by 
implementing the recent proposal of imaginary-valued mass difference to bypass the sign problem in lattice Monte Carlo calculations. 
The fully non-perturbative results thus obtained are analytically continued to real mass imbalance to yield the physical equation of state, 
providing predictions for upcoming experiments with mass-imbalanced atomic Fermi gases. 
In addition, we present an exact relation for the rate of change of the equation of state at small mass imbalances, showing that it is 
fully determined by the energy of the mass-balanced system.
\end{abstract}

\maketitle
%%%%%%%%%%%%%%%%%%%%%%%%%%%%%%%%%%%%%%%%%%%%%%%%%%%%%%%%%%

{\it Introduction.--~} Experiments with ultracold Fermi gases continue to move forward in their study 
of progressively richer quantum many-body systems, advancing our understanding of strongly interacting
matter at the interface of nuclear, atomic and condensed matter physics.
From the pioneering experiments realizing a condensate and determining the energy of unpolarized atomic clouds~\cite{FirstExperiments}, research has moved on in less than a decade to experiments varying temperature and polarization~\cite{PolarizedExp}, studies of Bose-Fermi mixtures~\cite{BoseFermiExp}, optical lattices~\cite{OpticalLatticeExp}, 
precise determinations of the equation of state~\cite{PrecisionExp},
and the list continues (see e.g. Ref.~\cite{ExpReview} {for reviews}). 
Now {many new} experiments with mixtures of a variety of different fermion species (such as ${}^{6}$Li, ${}^{40}$K, ${}^{161}$Dy, ${}^{163}$Dy, ${}^{167}$Er)
{appear to be possible
in the near future (see e.g. Ref.~\cite{GrimmPC}),} giving us an unprecedented opportunity to 
understand the effects of mass imbalance in strongly coupled gases.
In particular, it is possible to achieve mixtures with mass imbalances smaller than the one associated
with a ${}^{6}$Li-${}^{40}$K-mixture for which three-body effects are already expected to be significant~\cite{BraatenHammer}, thus making 
experimental studies more challenging.

Of the wide variety of regimes explored in ultracold fermions, one has received unparalleled attention. This is the
{so-called ``unitary'' Fermi gas,} which is realized by adjusting an external magnetic field on a dilute system of
two fermion species, setting the system close to a broad Feshbach resonance. 
{Since} the interaction range $r_s^{}$ is effectively 
zero relative to the interparticle spacing $n^{-1/3}$ {in that limit} and the $s$-wave scattering length $a_s^{}$ is very large 
(on the order of hundreds to thousands of Bohr radii), i.e. $r_s^{} \ll n^{-1/3} \ll a_s^{}$, 
{the system is scale-free} (except for $n$, as in a non-interacting gas). This disappearance of dynamical scales 
at unitarity has brought about interest from the nuclear physics area (where the unitary gas is a model for neutron matter~\cite{Bertsch} and 
has been proposed as a starting point for perturbative nuclear structure calculations~\cite{KSW}). Indeed, the strong pairing 
displayed at (and around) resonance results in a relatively high critical temperature ($T_c^{}\simeq 0.15-0.17$, in units of the Fermi 
energy~\cite{UnitaryTc}) with observed pairing correlations above $T_c^{}$ \cite{ExpPseudoGapUnitary}, as in high-$T_c^{}$ superconductors. 
In addition, the realization that the physics of the unitary limit is directly connected to a non-relativistic conformal fixed point~\cite{NishidaSon,Enss}, 
has spurred interest from the string-theory side, in connection with gauge-gravity duality~\cite{NRConformalStuff}.

In this work, we determine the zero-temperature equation of state of a mass-imbalanced unitary Fermi gas, by computing 
the ground-state energy as a function of the mass imbalance. As three-body effects become significant for large mass imbalances,
we focus on the equation of state for mass imbalances below the one
associated with a ${}^{6}$Li-${}^{40}$K-mixture. A natural question
for imbalanced systems of this kind is whether they undergo a 
quantum phase transition at a critical imbalance. This work is a first step towards answering this question in
a controlled and non-perturbative way. To our knowledge, there are no previous {\it ab initio}
calculations covering the range of mass imbalances presented here.

Because the unitary regime is strongly coupled, and there is no small parameter 
to formulate an expansion, computational methods are needed {for quantitative predictions}. 
For a many-body problem of this kind, Monte Carlo methods are 
the tools of choice. However, the presence of a finite mass-imbalance introduces a ``sign problem" 
as in systems with finite polarization, or in QCD in the presence of finite baryon density~\cite{SignProblemQCD}. 
This is a serious roadblock in many fields (condensed matter, atomic physics, high- and low-energy nuclear physics).
For mass-imbalanced Fermi gases, this implies that our present understanding is based
on mean-field studies partly amended to account for fluctuation effects to some extent, 
see, e.g., Ref.~\cite{mfstudies} for early ground-breaking studies.
While such studies give qualitative access to the features of these systems, it is also
known that mean-field theory does not predict the energy of the mass-balanced unitary Fermi gas correctly.
In fact, the mean-field result (see, e.g., Refs.~\cite{revchevy} for reviews) for the ground-state energy is about 50\% larger than the
accepted values from Monte Carlo calculations~\cite{Carlson:2003zz}.

To make progress in spite of the sign problem, we implement a method we proposed recently in Refs.~\cite{Braun:2012ww, BraunMassImbalance}. 
Borrowing ideas from a technique originally devised for lattice QCD~\cite{SignProblemQCD}, we introduced an imaginary mass imbalance,
such that each species has a complex mass, but one is the complex conjugate of the other. As a result,
the sign problem is avoided. However, the data now needs to be analytically continued to real mass imbalance.
This can be done in many ways, but the imaginary-mass calculations are performed in a fully non-perturbative 
fashion without ambiguities. It should be noted that this technique has never been applied to non-relativistic systems before.
Moreover, because the system we study is simple yet strongly coupled, our calculations may also shed light {on aspects of similar
methods on the lattice QCD side.} 
 
{\it Computational technique.--~}
Following the notation of Ref.~\cite{BraunMassImbalance}, the Hamiltonian~$\hat{H}$ of two Fermi species, \mbox{$\uparrow$ and $\downarrow$}, 
with a zero-range interaction is
\be
\hat{H}\!=\! 
\int\!d^{3}x\! \left (
 \sum_{\sigma=\uparrow,\downarrow}
  \hat{\psi}_{\sigma}^{\dagger}(\mathbf{x})\left(\frac{-\vec{\nabla}^2}{2m_{\sigma}}\right)\hat{\psi}_{\sigma} (\mathbf{x})
  + \bar{g} \hat{\rho}_\uparrow (\mathbf{x}) \hat{\rho}_\downarrow (\mathbf{x}) \right )\nn
\ee
and can be viewed as the sum
of the kinetic operators~$\hat{T}_{\uparrow,\downarrow}$ 
associated with the two fermion species and an operator~$\hat{V}$ specifying the interaction,~$\hat{H}=\hat{T}_{\uparrow}+\hat{T}_{\downarrow}+\hat{V}$.
The operators~$\hat{\rho}^{}_{\uparrow,\downarrow}$ are the particle-density operators,
and the masses of the species are~$m_{\uparrow}$ and~$m_{\downarrow}$. 
 To simplify the discussion of mass imbalanced systems, we define a dimensionless imbalance 
parameter~$\bar{m}=(m_{\downarrow}-m_{\uparrow})/(m_{\downarrow}+m_{\uparrow})$,
which maps the problem to a finite interval, $0\!\leq\! |\bar{m}|\! < \!1$.
For convenience, we shall take units such that $\hbar = k_B^{}=1$,
as well as~$m_0=(m_{\uparrow}+m_{\downarrow})/2=1$. With these conventions,
we {have~$\bar{m}\approx 0.74$ for} a ${}^6$Li-${}^{40}$K mixture.

We implement a projection quantum Monte Carlo (QMC) algorithm (see e.g. Ref.~\cite{Drut:2012md})
on a spacetime lattice, whereby we start with a Slater determinant $|\psi_0 \rangle$ as a guess for the ground-state
wavefunction and project towards the ground state by evolving in imaginary time. This is accomplished by applying the transfer matrix 
$\mathcal T = \exp{(-\tau \hat H)}$, which we factorize in the Trotter-Suzuki fashion. 
A Hubbard-Stratonovich transformation is then used to represent the interaction, allowing
us to write the transfer matrix as (see e.g. Ref.~\cite{Drut:2012md} for further details):
\beq
\label{T_HS1}
\mathcal T = \int \mathcal D \sigma\ \mathcal T^{}_{\uparrow}[ \sigma] \mathcal T^{}_{\downarrow}[ \sigma]\,,
\eeq
where $\mathcal D \sigma = \prod_{\bm{i}} d\sigma_{\bm{i}}/(2\pi)$, and 
$\sigma_{\bm{i}}$ is an external auxiliary field taking values between $-\pi$ and $\pi$ at each point $i$ in the spatial lattice.
The partition sum (or rather its zero-temperature analogue) is then given by
\beq
\label{Z}
\mathcal Z^{}_0(\beta) \equiv  \langle \psi_0 | \prod^{N_\tau}_{t=1} \mathcal T | \psi_0 \rangle
= \int \mathcal D \sigma\ \det \left[ \prod^{N_\tau}_{t=1} \mathcal T^{}_{\uparrow}[ \sigma] \mathcal T^{}_{\downarrow}[ \sigma] \right]\,.
\eeq
Here, $\sigma$ is to be regarded as defined on a spacetime lattice of $N_x^3 \times N^{}_\tau$ points,
with lattice spacing $\ell=1$ (by definition) in the spatial directions and $\tau$ in the {time direction,~$\beta=N_{\tau}\tau$.}
The determinant is taken over the single-particle space of the orbitals that make up {the $N$-fermion state $| \psi_0 \rangle$.}

In Eq.~\eqref{Z}, we observe that 
the fermion determinant factorizes into a determinant for each fermion species. In the absence of mass imbalance, 
and if the interaction is purely attractive ($\bar{g}>0$ in our convention), these determinants are real and identical. The product is thus positive semidefinite and can therefore be used as a probability measure in a QMC calculation. In the presence of a finite mass imbalance,
on the other hand, this is no longer the case, which spoils the naive application of QMC methods. By using an imaginary 
mass imbalance, however, the determinants become complex conjugates of each other,
which again makes their product positive semi-definite and therefore amenable to standard QMC 
methods~\cite{BraunMassImbalance}. In finite temperature calculations, it is in principle also possible to calculate at finite 
spin imbalance via imaginary chemical potential differences~\cite{Braun:2012ww}.

To connect the bare lattice theory to the physical parameters (namely the scattering length $a_s^{}$), we
utilize L\"uscher's formula~\cite{Luescher} for the relation between the phase shift and the energy eigenvalues 
of the two-body problem in a box. As the lattice eigenvalues depend directly on the bare coupling $\bar{g}$ and the reduced mass
$m_r^{} = (m_0/2)(1 - {\bar m}^2)$, it is a simple matter to tune these to match the desired physics, 
see, e.g., Refs.~\cite{box,Drut} for details. This results in $\bar{g}({\bar m}) = \bar{g}({\bar m}\!=\!0)/(1 - {\bar m}^2)$.

{\it Data Analysis.--~} 
In the present work, we focus on the (dimensionless) equation of state~$\xi(\bar{m})$, which 
is defined as the ratio of the energy of the interacting $N$-body problem evaluated
at a given~$\bar{m}$, to the energy of the corresponding non-interacting mass-balanced Fermi gas. For~$\bar{m}=0$,
this definition matches the one of the so-called {\it Bertsch} parameter. 
To be more specific, we calculated the $\beta \epsilon_{\rm F}$-dependence of systems of
many particles in cubic lattices of different sizes (see Table~\ref{Table:FitParameters2}), 
and for imaginary-mass imbalances in the range $i\bar{m}\!=\! 0,0.025,0.05,\dots,\!1$.
Here,~$\epsilon_{\rm F}$ denotes the Fermi energy of the non-interacting gas at the same density.
The data was then extrapolated to large $\beta \epsilon_F$ in a standard fashion~\cite{Drut}.\footnote{For the lattices and particle 
numbers studied here, we have found that the analytic continuation could be performed
before or after the extrapolation to $\beta \epsilon_{\rm F} \to \infty$ without significant change in the results.}

For each set of parameter values ($N^{}_x$, $N$, $\beta \epsilon_F$, $\bar{m}$) we used approximately 
500 decorrelated samples of the auxiliary field, which yields a statistical uncertainty of the order of $5\%$. The lattice length and energy 
units were set by the spatial lattice spacing $\ell= 1$, and the imaginary-time spacing $\tau = 0.05$ (in the units determined by~$\ell$). 
To perform the analytic continuation from imaginary to real $\bar{m}$, we employed an ansatz
for the form of the equation of state as a function of~$\bar{m}$, to which we fit the data.

{\it Results.--~} 
Naturally, an analytic continuation of the data is only meaningful if the partition sum~$\mathcal Z_0$
is an analytic function of~$\bar{m}$ in a finite domain about~$\bar{m}=0$. 
In practice, however, little is known about the analytic properties of the full partition sum.
Therefore any analytic insight is important to guide the continuation.
A very first understanding of the~$\bar{m}$-dependence of the equation of state~$\xi$ can be obtained from the 
free mass-imbalanced Fermi gas. In that case,~$\xi_{\text{free}}\equiv E_{\text{free}}(\bar{m})/E_{\text{free}}(0)=1/(1-\bar{m}^2)$
for real-valued~$\bar{m}$. Note that the partition sum is invariant
under~$\bar{m}\to -\bar{m}$.

To better understand the effect of fermion interactions on the functional form of~$\xi(\bar{m})$,
we consider a mean-field analysis.
{In Refs.~\cite{BraunMassImbalance,Braun:2014ewa}, it was} found that
the radius of convergence~$r_{\bar{m}}$ associated with the~$\bar{m}$-dependence of the partition sum is
maximal for the {unpolarized} system,~$r_{\bar{m}}=1$. For the mean-field equation of state as a function of (real-valued) 
mass-imbalances, we obtain~$\xi_{\rm mf}(\bar{m})=\xi_{\rm mf}(\bar{m}\!=\!0)/(1-\bar{m}^2)$,
where~$\xi_{\rm mf}(\bar{m}\!=\! 0)\approx 0.6$. Thus, the conventional {\it Bertsch} parameter~$\xi_{\rm mf}(\bar{m}\!=\!0)$ completely
determines the coefficients of an expansion in powers of~$\bar{m}^2$. 
The analyticity of the mean-field equation of state in the range~$0\leq \bar{m}<1$
can be understood from the fact that the system does not undergo a (quantum) phase transition from the superfluid phase to a normal phase as function
of~$\bar{m}$ for~$N_{\uparrow}=N_{\downarrow}$ (see e.g., Ref.~\cite{Braun:2014ewa}). Indeed, the existence of a phase transition 
at a given critical value~$\bar{m}_{\text{cr}}$ would be associated with a non-analytic behavior of $\mathcal Z_0$, 
and therefore with a non-analytic behavior of the observables. 

Beyond the mean-field approximation, an analysis of the analytic properties of the full partition sum is difficult. From the path-integral representation
Eq.~\eqref{Z}, however, it is a simple matter to derive an exact differential equation for the equation of state:
\beq
\frac{\partial \langle \hat{H}\rangle}{\partial \bar{m}^2}= \frac{1}{1-\bar{m}^2}\left(  \langle \hat{H}\rangle
- \langle\hat{T}_{\Sigma}\rangle\right)\,,\label{eq:de}
\eeq
where~$\langle\hat{T}_{\Sigma}\rangle = \frac{1}{2} \left(\langle \hat{T}_{\uparrow}+\hat{T}_{\downarrow}\rangle 
- \frac{1}{\bar{m}} \langle \hat{T}_{\uparrow} - \hat{T}_{\downarrow}\rangle\right)$, and the mass 
derivative is taken along the line of constant physics, i.e. such that 
the two-body scattering parameters remain constant.\footnote{
{The proof of Eq.~\eqref{eq:de} makes use of 
the {\it Hellmann-Feynman} theorem, from which it follows 
that~$\partial \langle \hat{H}\rangle/\partial \bar{m}^2 = \langle \partial\hat{H}/\partial\bar{m}^2\rangle$.}}
We may interpret the quantity~$\langle\hat{T}_{\Sigma}\rangle$ as a measure of the difference between the kinetic energies of the spin-up and down
fermions in the presence of an interaction. Indeed,~$\langle\hat{T}_{\Sigma}\rangle =0$ for the free gas for all~$\bar{m}$, as well as for the 
interacting mass-balanced system.
The initial condition for the differential equation~\eqref{eq:de} is given
by the energy of the mass-balanced system, and the dimensionless equation of state~$\xi$ is given 
by~$\xi(\bar{m})=\langle \hat{H}\rangle(\bar{m})/E_{\text{free}}(\bar{m}=0)$.
Note also that the mean-field equation of state 
is obtained from the differential equation~\eqref{eq:de} by setting~$\langle\hat{T}_{\Sigma}\rangle\to0$.
\begin{table}[t]
\begin{center}
\begin{tabular}{|c||c|c|c|}
\hline\hline
$N/N_x^3$ & $\xi(\bar{m}\!=\!0)$ & $\xi^{(1)}$ & $\chi^2/\text{dof}$ \\
\hline\hline 
$0.05$ & $0.414\pm 0.012$ & $0.461\pm 0.181$ & $0.5$\\ 
\hline\hline
\end{tabular}
\caption{\label{Table:FitParameters1}
Estimates for the {\it Bertsch} parameter~$\xi(\bar{m}\!=\!0)$ and the curvature~$\xi^{(1)}:=d\xi(\bar{m})/d\bar{m}^2|_{\bar{m}=0}$ 
for real mass imbalance in the {large-volume} limit, fitting with $\bar{m}\!\leq\!0.4$.}
\end{center}
\end{table}
\begin{table}[t]
\begin{center}
\begin{tabular}{|c|c||c|c|c|}
\hline\hline
$N_x$ & $N$ & $\xi(\bar{m}\!=\!0)$ & $\xi_{M}$ & $\chi^2/\text{dof}$ \\
\hline\hline 
$8$ & $24$ & $0.449\pm 0.002$ & $0.496\pm 0.010$ & 0.9  \\ 
$10$ & $46$ & $0.431\pm 0.002$ & $0.631\pm 0.011$ & $2.0$ \\ 
$12$ & $80$ & $0.444\pm 0.002$ & $0.532\pm 0.010$ & $0.4$\\ 
$\infty$ & --- & $0.420\pm 0.007$ & $0.693\pm 0.111$ & $0.3$\\
\hline\hline
\end{tabular}
\caption{\label{Table:FitParameters2}
Parameters for the global fit function given in Eq.~\eqref{eq:fitf} for a fixed density $N/N_x^3\approx 0.05$ as 
obtained from three different particle {numbers $N = N_{\uparrow}+N_{\downarrow}$ ($N_{\uparrow}^{}=N_{\downarrow}^{}$)}
and volumes $N_x^3$. The error bars of these fits
result from statistical and large-$\beta\epsilon_{\rm F}$-extrapolation errors. Estimates 
in the {large-volume} limit ($N_x\to\infty$ with fixed $N/N_x^3\approx 0.05$) are also shown,
{where a $1/N_x$-extrapolation of the data was performed before the fit.}}
\end{center}
\end{table}
Our analytic insight into the $\bar{m}$-dependence of the equation of state suggests that it may be insufficient to fit our QMC data to a 
low-order truncation of a polynomial in~$\bar{m}^2$. Below, we therefore
only fit our data for small~$\bar{m}$ to a polynomial in~$\bar{m}^2$ to show 
that our numerical data obeys~$\partial \langle \hat{H}\rangle/\partial \bar{m}^2|_{\bar{m}=0}= \langle \hat{H}\rangle|_{\bar{m}=0}$, which follows
from Eq.~\eqref{eq:de} and $\langle\hat{T}_{\Sigma}\rangle=0$ at~$\bar{m}=0$. To provide a global description of our 
QMC data for imaginary-valued mass imbalances, we employ a Pad\'{e} approximant:
\be
\xi(\bar{m})=\frac{\xi(\bar{m}\!=\!0)}{1+\xi_{M}\bar{m}^2}\,,
\label{eq:fitf}
\ee
where~$\xi(\bar{m}\!=\!0)$ and~$\xi_{M}$ are the only two fit parameters. Note that this ansatz violates the  
constraint~$\partial \langle \hat{H}\rangle/\partial \bar{m}^2|_{\bar{m}=0}= \langle \hat{H}\rangle|_{\bar{m}=0}$
if~$\xi_{M}\neq 1$. While we could consider more sophisticated approximants to include this constraint, we have found that our 
present ansatz already provides a reasonable parameterization for the equation of state for~$0\leq |\bar{m}|<1$ (see below).

\begin{figure}[t]
\includegraphics[width=0.97\columnwidth]{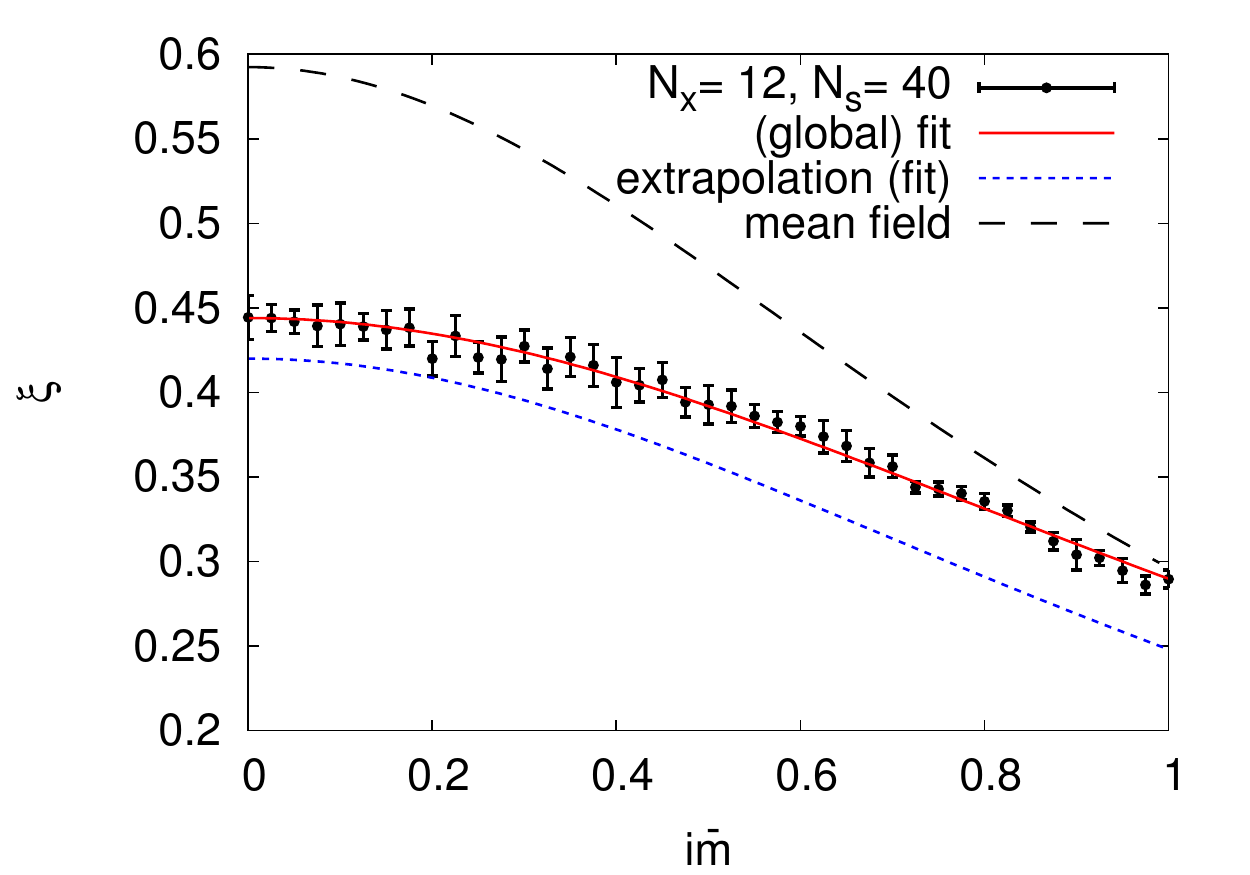}
\caption{\label{Fig:EoSI}(color online) Equation of state, with statistical error bars, and corresponding fit function 
(see Eq.~\eqref{eq:fitf}) {for~$N_x=12$ and~$N=80$ 
($N/N_x^3\approx 0.05$)} as a function of imaginary mass imbalance ${\rm i}\bar{m}$;
see Table~\ref{Table:FitParameters2} for the
fit parameters. Our estimate {in} the {large-volume} limit (extrapolation) and the result 
from mean-field theory are also shown.}
\end{figure}
\begin{figure}[t]
\includegraphics[width=0.97\columnwidth]{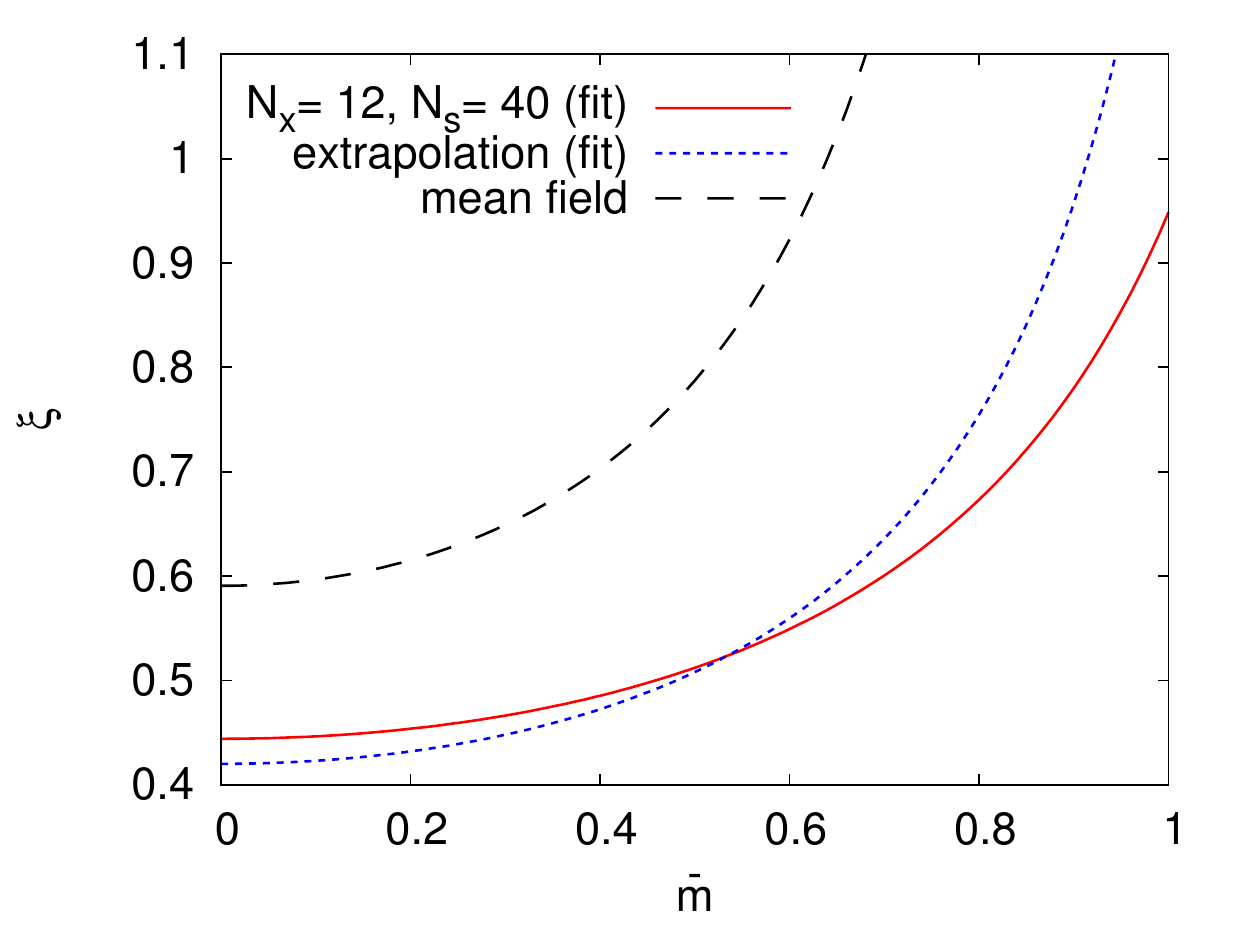}
\caption{\label{Fig:EoSR}(color online) 
Equation of state as obtained from analytic continuation of the fit function~\eqref{eq:fitf} (see Table~\ref{Table:FitParameters2} for the fit parameters) 
as a function of real mass imbalance~$\bar{m}$. Our estimate {in} the {large-volume} limit (extrapolation) and the result 
from mean-field theory are also shown.}
\end{figure}
We next discuss our QMC results for~$\xi(\bar{m})$, beginning with small values of~$\bar{m}$. 
Our analytic calculation predicts that~$\partial \langle \hat{H}\rangle/\partial \bar{m}^2|_{\bar{m}=0}= \langle \hat{H}\rangle|_{\bar{m}=0}$, i.e. the
curvature of the equation of state at~$\bar{m}=0$ is fully determined by the {\it Bertsch} parameter~$\xi(0)$. 
Our QMC data for imaginary mass imbalances agrees with this statement as seen from a
fit to the ansatz~$\xi(\bar{m})=\xi(\bar{m}\!=\!0) - \xi^{(1)}\bar{m}^2$, see Table~\ref{Table:FitParameters1}.
For the fit, we have only used data for~$\bar{m}\leq 0.4$ as obtained 
from a {$1/N_x$-extrapolation of the original QMC data for~$N_x=8,10,12$}
to the infinite-volume limit while keeping the density~$N/N_x^3\approx 0.05$ fixed. {We chose this (relatively low) density to
(at least partially) avoid finite-range effects. The continuum limit can be reached by studying the low-density limit.}
Already the data for the smallest lattice size, $N_x=8$, agrees within error bars with our analytic prediction: 
we obtain~$\xi(\bar{m}\!=\!0)\approx 0.455\pm 0.005$ and~$\xi^{(1)}\approx 0.603\pm 0.228$. Note
that our estimate for the {\it Bertsch} parameter in the {large-volume} limit agrees with previous QMC 
studies~\cite{Carlson:2003zz}, up to finite-range effects (see Table~\ref{Table:FitParameters1}).

In Figs.~\ref{Fig:EoSI} and~\ref{Fig:EoSR}, we show the equation of state
as a function of imaginary and real mass imbalance, respectively, as obtained from an analytic continuation based on Eq.~\eqref{eq:fitf}.
We observe that mean-field theory significantly overestimates the ground-state energy, which 
is well-known for the mass-balanced case~\cite{Carlson:2003zz}.
Increasing the mass imbalance, the mean-field equation of state shows a much stronger dependence 
on~$\bar{m}$ than our QMC results. For small mass imbalance, this already {follows from Eq.~\eqref{eq:de}.} For large real-valued mass 
imbalances, this can be deduced from Fig.~\ref{Fig:EoSR}.
For example, {for~$\bar{m}\approx 0.74$} (${}^{6}$Li-${}^{40}$K-mixture), 
we find~$\xi_{\text{mf}}(\bar{m})/\xi_{\text{QMC}}(\bar{m})\approx 2.0$ when compared to our estimate for the equation
of state in the {large-volume} limit. Note
that~$\xi_{\text{mf}}\to \infty$ for~$\bar{m}\to 1$, whereas the analytic continuation of the lattice data suggests that~$\xi(\bar{m})$ remains 
finite in this limit. This should be taken with some care, as our results for large real-valued mass imbalances depend strongly on the 
details of the ansatz for the fit function underlying the analytic continuation. However, we have checked that even with more sophisticated fit functions 
(Pad\'{e} approximants with up to four parameters), the uncertainty for~$\xi(\bar{m})$ {at~$\bar{m}\approx 0.74$} is about $30\%$ at most and decreases 
rapidly for decreasing mass imbalances. A detailed analysis of~$\xi(\bar{m})$ for various densities and lattices is beyond the present work 
and will be presented elsewhere~\cite{DBR}. {For the time being, 
we only would like to note that our (present) results are qualitatively in agreement with 
previous direct QMC calculation for~{$\bar{m}\approx 0.74$} (${}^6$Li-${}^{40}$K mixture)~\cite{Gezerlis},
both indicating that mean-field theory significantly overestimates the ground-state energy.}

{\it Summary and Conclusions.--~} We have presented a first lattice MC determination of the
equation of state of resonantly interacting fermions with finite mass imbalance. This \emph{ab initio},
fully non-perturbative calculation was accomplished by implementing our recent proposal of
taking the mass imbalance to the imaginary axis, where we can calculate without a sign problem. 
The data thus obtained was analytically continued via a simple ansatz to real mass imbalances.
To simplify this first attempt, and provide a useful benchmark, we focused on
the case of equal particle numbers $N_\uparrow=N_\downarrow$, such that the Fermi momenta coincide, and
therefore a (quantum) phase transition at finite $\bar m$ is not expected. 

Although we do not aim for high accuracy in this first study, our analysis indicates that mean-field 
studies not only significantly overestimate the ground-state energy, but also its 
change when the mass imbalance is increased.
This observation is already of great importance, as many of the predictions for mass-imbalanced ultracold gases rely on the 
mean-field approximation. Future experiments will open up the possibility to measure at least parts of the equation of state 
for~$\bar{m}\lesssim 0.7$. Our present calculation of the latter and, in particular, of the curvature
can then be tested directly.

Our work verifies that the method of imaginary mass imbalances is feasible. It should now be possible to furnish a number of predictions 
for an upcoming set of ultracold-atom experiments, paving the way for future calculations
with mass imbalance in a variety of systems and situations, e.g. away from unitarity, in various dimensions, 
at finite temperatures, in an external potential, and including both imaginary mass and polarization.

%%%%%%%%%%%%%%%%%%%%%%%%%%%%%%%%%%%%%%%%%%%%%%%%%%%%%%%%
{\it Acknowledgments.--~} J.B. and D.R. acknowledge support by the DFG under Grant BR 4005/2-1 and
by HIC for FAIR within the LOEWE program of the State of Hesse. Moreover, JB acknowledges useful
discussions {with R.~Grimm and H.-W.~Hammer.}
This work was supported in part by the U.S. National Science Foundation under Grant No. PHY{1306520}.

%%%%%%%%%%%%%%%%%%%%%%%%%%%%%%%%%%%%%%%%%%%%%%%%%%%%%%%%
\bibliographystyle{h-physrev3}

\end{document}